\documentclass[12pt]{article}
\usepackage{amssymb}
\usepackage{graphicx}
\usepackage{amsmath}

\begin{document}

\begin{titlepage}
\begin{flushright}
PUPT-1939

hep-th/0006132
\end{flushright}

\vspace{7 mm}

\begin{center}
{\huge String theory as a universal language}
\end{center}
\vspace{10 mm}
\begin{center}
{\large  

A.M.~Polyakov\\
}
\vspace{3mm}
Joseph Henry Laboratories\\
Princeton University\\
Princeton, New Jersey 08544
\end{center}
\vspace{7mm}
\begin{center}
{\large Abstract}
\end{center}
\noindent This article, based on the Klein lecture, contains some new results and new speculations on various topics. They include discussion of  open strings in the AdS space, unusual features of D-branes,
conformal gauge theories in higher dimensions. We also comment on the infrared screening of the  cosmological constant
and on the "brane worlds"
\vspace{7mm}
\begin{flushleft}
June 2000

\end{flushleft}
\end{titlepage}
%

\section{\protect\bigskip Introduction}

String theory is a beautiful and dangerous subject. On one hand it is a top
achievement of theoretical physics exploiting the most advanced and daring
methods. On the other --- without a guidance from the experiment it can easily
degenerate into a collection of baroque curiosities, some kind of modern
alchemy looking for philosopher's stone.

This danger can be somewhat reduced if we try to study string theory in
connection with some concrete physical problem and then extrapolate the
gained experience to the Planck domain unreachable by experiments. This is a
well established strategy in theoretical physics. For example one can learn
about the Cherenkov radiation while studying supersonic aerodynamics. And
usually there is the ''back-reaction'': the technical progress at the
frontier turns out to be helpful in solving the old problems. Thus, it is
conceivable that string theory will provide us with the language for the
future theoretical physics.

In this lecture I will examine a number of problems in which the language of
string theory is appropriate and effective. We begin with the problem of
quark confinement. The task here is to find the string description of the
color-electric flux lines emerging in QCD. Recently there has been a
considerable progress in this field. Various aspects of it have been
reviewed in [1, 2]. I shall not review again these developments and instead
concentrate on new results. After that we will discuss some general features
of D-branes, conformal gauge theories in higher dimensions, and speculations
concerning the cosmological constant.

\section{The image of gluons and zigzag symmetry}

As was explained in the above references, a string theory, needed to
describe gauge fields in 4 dimensions, must be formulated in the 5d space
with the metric 
\begin{equation}
ds^{2}=d\varphi ^{2}+a^{2}(\varphi )d\vec{x}^{2}
\end{equation}

This curved 5d space is a natural habitat for the color-electric flux
lines. If apart from the pure gauge fields there are some matter fields in
the theory, they must correspond to extra degrees of freedom on the world
sheet. In some cases these extra degrees of freedom can be balanced so that
the field theory $\beta -$function is equal to zero. These cases are the
easiest ones, since the conformal symmetry of the field theory requires
conformal symmetry of the string background and determines it completely:
$$
a(\varphi )\sim e^{\alpha \varphi },
$$
where $\alpha $ is some constant. After an obvious change of variables the
metric takes the form 
\begin{equation}
ds^{2}=\sqrt{\lambda }y^{-2}(dy^{2}+d\vec{x}^{2}),
\end{equation}
where $\lambda $ is related to the coupling constant of field theory. When $%
\lambda \gg 1$ the curvature of this 5d space is small and the 2d sigma
model describing the string in the above background is weakly coupled. This
greatly simplifies the analyses and we will concentrate on this case. Our
aim in this section will be to demonstrate that open strings in this
background have some very unusual properties, allowing to identify them with
gluons.

Before starting let us recall that at present we have two possible
approaches to the question of field-string correspondence. None of them is
fully justified but both have certain heuristic power. In the first approach
one begins with a stack of D-branes describing a gauge theory and then
replace the stack by its gravitational background. In the second approach
one doesn't introduce D-branes and starts directly with the sigma model
action, adjusting the background so that the boundary states of this string
describe the gauge theory. The key principle here is the zigzag symmetry.
This is a requirement that these boundary states consist of vector gluons
(and matter fields, if present) and nothing else.

The situation is very unusual. Normally we have an infinite tower of states
in both open and closed string sectors. Here we need a string theory in
which the closed string sector contains an infinite number of states, while
the open sector has a finite number of the field-theoretic states. Our first
task will be to explore how this is possible.

To set the stage, let us remember how open strings are treated in the
standard case [3]. One begins with the action 
\begin{equation}
S=\frac{1}{2}\int_{D}(\partial x)^{2}+i\int_{\partial D}A_{\mu }dx_{\mu },
\end{equation}
where $D$ is a unit disc, $\partial D$ --- its boundary, and $A_{\mu }$ is a
vector condensate of the open string states. The possible fields $A_{\mu }$
are determined from the condition that the functional integral 
\begin{equation}
Z[A]=\int Dx\,e^{-S}
\end{equation}
is conformally invariant. The explicit form of this condition is derived by
splitting 
\begin{equation}
x=c+z,
\end{equation}
where $c$ is a slow variable, while $z$ is fast, and integrating out $z$.
Conformal invariance requires vanishing of the divergent counterterms and
that restricts the background fields. It is convenient to integrate first
the fields inside the disk with the fields at the boundary being fixed. That
gives the standard boundary action 
\begin{equation}
S_{B}=\frac{1}{2}\int \frac{dudv}{(u-v)^{2}}(x(u)-x(v))^{2}+i\int A_{\mu
}dx_{\mu }.
\end{equation}
We see that $x_{\mu }$ are the Gaussian fields with the correlation
function (in the momentum space) 
\begin{equation}
<x_{\mu }(p)x_{\nu }(-p)>\sim \delta _{\mu \nu }\mid p\mid ^{-1}.
\end{equation}
Expanding the second term in $z$ we obtain 
\begin{equation}
S_{B}\approx \frac{1}{2}\sum \mid p\mid z(p)z(-p)+\int \nabla _{\lambda
}F_{\sigma \mu }(c)\frac{dc_{\mu }}{ds}z_{\lambda }(s)z_{\sigma }(s)ds.
\end{equation}
Using the fact that 
\begin{equation}
<z_{\lambda }(s)z_{\sigma }(s)>\sim \delta _{\lambda \sigma }\int^{\Lambda }%
\frac{dp}{\mid p\mid },
\end{equation}
where $\Lambda $ is an ultraviolet cut-off, we obtain as condition that the
divergence cancel (in this approximation) 
\begin{equation}
\nabla _{\lambda }F_{\lambda \mu }=0.
\end{equation}

This is the on-shell condition for the massless string mode. Qualitatively
the same treatment is applicable to the massive states as well. In this case
one perturbs $S_{B\text{ }}$ with the operator 
\begin{equation}
\Delta S_{B}=\int ds\Psi (x(s))(\overset{.}{x}^{2}(s)h^{-2}(s))^{n}h(s),
\end{equation}
where $\Psi $ is the scalar massive mode at the level $2n$ and $h(s)$ is the
boundary metric on the world sheet needed for the general covariance of this
expression. Conformal invariance of this perturbation means that the $h(s)$
dependence must cancel. The cancellation occurs between the explicit $h$
-dependence in the above formula and the factors coming from the quantum
fluctuations of $x(s)$. These factors appear because in the covariant theory
the cut-off is always accompanied by the boundary metric 
\begin{equation}
h(s)(\Delta s)_{\min }^{2}=a^{2},
\end{equation}
\begin{equation}
\Lambda ^{2}=\frac{1}{(\Delta s)_{\min }^{2}}=\frac{1}{a^{2}}h(s),
\end{equation}
where $a$ is an invariant cut-off.

In the one loop approximation (which is not, strictly speaking, applicable
here, but gives a correct qualitative picture) we have 
\begin{equation}
\partial ^{2}\Psi -M_{n}^{2}\Psi =0,
\end{equation}
\begin{equation}
M_{n}^{2}\sim n.
\end{equation}

This is the on-shell condition for the massive string mode and it was
obtained, let us stress it again, from the cancellation between the
classical and quantum $h(s)$ dependence

Now we are ready to attack the AdS case. Let us consider the string action
in this background 
\begin{equation}
S=\sqrt{\lambda }\int_{D}\frac{(\partial x_{\mu })^{2}+(\partial y)^{2}}{%
y^{2}}+...,
\end{equation}
where we dropped all fermionic and RR terms. This is legitimate in the WKB
limit $\lambda \gg 1$ which we will study in this section. To find the
counterterms we must once again calculate the boundary action and $%
<z_{\lambda }(s)z_{\mu }(s)>$. It is not as easy as in the previous case,
but this well-defined mathematical problem was solved in [4, 5]. The answer
has the following form 
\begin{equation}
S_{cl}=\sqrt{\lambda }\int ds_{1}ds_{2}\kappa _{\mu \nu }(s_{1},s_{2})z_{\mu
}(s_{1})z_{\nu }(s_{2}).
\end{equation}

After introducing variables $s=\frac{s_{1}+s_{2}}{2}$ and $\sigma
=s_{1}-s_{2}$ and taking the Fourier transform with respect to $\sigma $ we
obtain the following asymptotic for the kernel in the mixed representation 
\begin{equation}
\kappa _{\mu \nu }(p,s)\underset{p\rightarrow \infty }{\approx }\frac{\mid
p\mid ^{3}}{(c'(s))^{2}}[3\frac{c'_{\mu }c'_{\nu
}}{(c')^{2}}-\delta _{\mu \nu }].
\end{equation}
>From this it follows that 
\begin{equation}
<z_{\mu }(s)z_{\nu }(s)>\varpropto \int^{\infty }\frac{dp}{\mid p\mid ^{3}}%
<\infty.
\end{equation}

The remarkable feature of this answer is that it implies that \emph{there is
no quantum ultraviolet divergences on the world sheet.} Hence, if we add to
the action the background fields 
\begin{equation}
\Delta S\sim \int A_{\mu }dx_{\mu }+\int \Psi (x(s))(x'(s))^{2}(h(s))^{-1}ds+...
\end{equation}
and treat it in the one loop approximation, we come to the following
conclusions. First of all, as far as the $A$-term is concerned, it is finite
for any $A_{\mu }(x)$ and thus describes the off-shell gluons. This
situation is in the sharp contrast with the standard case in which conformal
invariance implied the on-shell condition.

Now let us examine the massive mode (11). The only quantum dependence on the
cut-off comes from 
\begin{equation}
<z'_{\mu }(s)z'_{\nu }(s)>\sim \int \frac{dp}{\mid
p\mid }
\end{equation}
As a result we obtain the following counterterm 
\begin{equation}
\Delta S\sim \int \Psi (x(s))(x'(s))^{2}(\frac{\log h(s)}{h(s)}%
)ds
\end{equation}

We see that the only way to keep the theory conformally invariant in this
approximation is to set $\Psi =0$. There is also a possibility that at some
fixed value of $\lambda $ the $h(s)$ dependence will go away. However it is
impossible to cancel it by a suitable on-shell condition. All this happens
because, due to (19), the kinetic energy for the $\Psi $-term is not
generated.

There is one more massless mode in the AdS string which requires a special
treatment. Let us examine 
\begin{equation}
\Delta S=\int \Phi (x(s))\partial _{\bot }y(s)\,ds,
\end{equation}
where $\partial _{\bot }$ is a normal derivative at the boundary of the
world sheet (which lies at infinity of the AdS space). In the more general
case of AdS$_{\text{p}}\times $S$_{\text{q}}$ we have also a perturbation 
\begin{equation}
\Delta S=\int \Phi ^{i}(x)n^{i}(s)\partial _{\bot }yds.
\end{equation}

Here we must remember that the string action is finite only if [6, 5] 
\begin{equation}
\left( \partial _{\bot }y\right) ^{2}=\left( x'\right) ^{2}.
\end{equation}

It is easy to see that when we substitute the decomposition (5) into this
formula we get the logarithmic divergence (21) once again. We come to the
conclusion that the above perturbation is not conformal and must not be
present at the boundary. However in the case of (24) there is also a
logarithmic term, coming from the fluctuations of $n^{i}(s)$. In the
presence of space-time supersymmetry these two divergences must cancel since
the masslessness of the scalar fields is protected by the SUSY. Otherwise,
keeping the scalar fields massless requires a special fine tuning of the
background. It would be interesting to clarify the corresponding mechanism.%
\newline

Another interesting problem for the future is the fate of the open string
tachyon in the AdS space. So far we assumed that it is excluded by the GSO
projection. But in the purely bosonic string it may lead to some
interesting effects via Sen mechanism [7].

We come to the following conclusion concerning the spectrum of the boundary
states in the AdS-like background. It consists of a few modes which would
have been massless in the flat case. The infinite tower of the massive
states can not reach the boundary. The above finite set of states must be
associated with the fields of the field theory under consideration.

The full justification of this assertion requires the analyses of the
Schwinger-Dyson equations of the Yang-Mills theory. It is still absent, and
we give some heuristic arguments instead. The loop equation, expressing the
Schwinger-Dyson equations in terms of loops has the form 
\begin{equation}
\widehat{L}(s)W(C)=W\ast W
\end{equation}

where $W(C)$ is the Wilson loop and $\widehat{L}$ is the loop laplacian and
the right hand side comes from the self-intersecting contours. In recent
papers [4, 5] we analyzed the action of the loop Laplacian in the AdS space.
It was shown that at least in the WKB approximation and in the four
-dimensional space-time we have a highly non-trivial relation 
\begin{equation}
\widehat{L}(s)\sim T_{\bot \Vert }(s)
\end{equation}

where $T_{\bot \Vert }(s)$ is a component of the world sheet energy-momentum
tensor at the boundary. When substituted inside the string functional
integral the energy momentum tensor receives contributions from the
degenerate metrics only. These metrics describe a pinched disk, that is two
discs joined at a point. The corresponding amplitude is saturated by the
allowed boundary operators inserted at this point. That gives the equation
(26 ) provided that the boundary operators of the string are the same as the
fields of the field theory. Much work is still needed to make this argument
completely precise.

\section{The D-brane picture}

An alternative way to understand gauge fields-strings duality is based on
the D-brane approach. It is less general than the sigma model approach
described above, but in the supersymmetric cases it provides us with an
attractive visual picture. The logic of this method is based on the
fundamental conjecture that D-branes can be described as some particular
solitons in the closed string sector. One of the strongest arguments in
favor of this conjecture is that both D-branes and solitons have the same
symmetries and are sources of the same RR fields [ 8]. The gauge fields
-strings dualities then follows from the D-branes - solitons duality in the
limit $\alpha ^{^{\prime }}\rightarrow 0$. On the D- brane side only the
massless gauge field modes of the open string survive in this limit. On the
string theory side we have a near horizon limit [1] of the soliton metric [9
] , given by (1) . These two theories must be equivalent, if the basic
D-brane conjecture is correct.

The connection with the sigma model approach of the previous section follows
from the following argument. First of all the closed string background is
the same in both cases. As for the open strings, we placed their ends at the
boundary of the AdS space where $a^{2}\left( \varphi \right) \rightarrow
\infty .$ That means that the effective slope of this strings behaves as 
\begin{equation}
\alpha _{open}^{^{\prime }}\sim a^{-2}(\varphi )\rightarrow 0
\end{equation}

and thus only the massless modes are present. We said that the D-brane
approach is less general, because in the non-supersymmetric cases there
could exist solitons with the required boundary behavior, which are not
describable by any combination of D-branes in the flat space.

This fact is related to another often overlooked subtlety. The 3-brane
soliton has a metric [9] 
\begin{eqnarray}
ds^{2} &=&H^{-\frac{1}{2}}(r)(dx)^{2}+H^{\frac{1}{2}}(r)(dy)^{2} \\
r^{2} &=&y^{2};\ \ H(r)=1+\frac{L^{4}}{r^{4}}
\end{eqnarray}

It is often assumed that this metric is an extremum of the action 
\begin{equation}
S=S_{bulk}+S_{BI}
\end{equation}

where the first term contains the modes of the closed string, while the
second is the Born -Infeld action localized on the brane. In the equations
of motion the second term will give a delta function of the transverse
coordinates.

Would it be the case, where the 3-brane is located? From (30) it is clear
that the singularity of the metric is located in the complex domain $%
r^{4}=-L^{4}.$ Let us try to understand the significance of this fact from
the string-theoretic point of view. Consider a string diagram describing the
D-brane world volume in the arbitrary order in $\lambda =g_{s}N.$ It is
represented by a disc with an arbitrary number of holes. At each boundary
one imposes the Dirichlet conditions for the transverse coordinates. An
important feature of this diagram is that it is finite. This follows from
the fact that the only source of divergences in string theory are tadpoles
and for 3-branes their contribution is proportional to the integral $\int 
\frac{d^{6}k}{k^{2}}$ where $k$ is the transverse momentum. This expression
is infrared finite (which is of course very well known). Thus D-branes in
the flat space are described by the well defined string amplitudes. But that
contradicts the common wisdom, that one must determine the background from
the action (31) , because the \emph{flat space is not a solution, }once the
Born-Infeld term is added. More over if we try to deform the flat space, the
above disc with holes will loose its conformal invariance.

Let us analyze this apparent paradox. It is related to the fact on a sphere
conformal invariance is equivalent to the absence of tadpoles since for any
(1,1) vertex operator we have $<V>_{sphere}=0$. However on a disc this is
not true, the conformal symmetry doesn't forbid the expectation values of
vertex operators. On a disc conformal symmetry and the absence of tadpoles
are two different conditions. Which one should we use?

If we denote the bulk couplings by $\lambda $ and the boundary couplings by $%
\mu $ we can construct three different objects, the bulk central charge $%
c\left( \lambda \right) ,$ the ''boundary entropy'' [10] $b\left( \lambda
,\mu \right) $ and the effective action generating the S-matrix, $S(\lambda
,\mu )=c\left( \lambda \right) +b\left( \lambda ,\mu \right) $ . To ensure
conformal invariance we must have 
\begin{eqnarray}
\frac{\partial c}{\partial \lambda } &=&0 \\
\frac{\partial b}{\partial \mu } &=&0
\end{eqnarray}

This does not coincide in general with the ''no tadpole condition'' 
\begin{equation}
\frac{\partial S}{\partial \lambda }=\frac{\partial S}{\partial \mu }=0
\end{equation}

In the case of 3-branes the paradox is resolved in an interesting way. The
metric (30) has a horizon at $r=0$. When we go to the Euclidean signature
the horizon, as usual, shrinks to a non-singular point. As a result we have
a metric which solves the equation ( 32) and has no trace of the D-brane
singularity in it! The paradox is pushed under the horizon.

The conclusion of this discussion is as following. We have two dual and
different descriptions of the D-brane amplitudes. In the first description
we calculate the amplitudes of a disc with holes \emph{in the flat space. }%
In this description it is simply inconsistent to introduce the background
fields generated by D-branes.

In the second description we forget about the D-branes and study a
non-singular closed string soliton. The D-brane conjecture implies that we
must get the same answers in these two cases. The situation is analogous to
the one we have in the sine-gordon theory, which admits two dual
descriptions, either in terms of solitons or in terms of elementary
fermions, but not both.

Let us touch briefly another consequence of these considerations. When
minimizing the action ( 31) one can find a solution which is singular on the
3-brane and is AdS-space outside of it [ 11]. These solutions are known to
''localize'' gravitons on the brane and are the basis of the popular
''brane-world'' scenarios. It is clear that for the string-theoretic branes
this is not a physical solution because the world volume does not contain
gravity (being described by the open strings). As we argued above, there
must be a horizon, not a singularity. Technically this happens because in
string theory the Born-Infeld action is corrected with the Einstein term $%
\sim \int R\sqrt{g}d^{p+1}x$ , coming from the finite thickness of the
brane. It can be shown that the coefficient in front of this term (which is
fully determined by string theory) is tuned so that the localization is
destroyed. There are no worlds on D-branes. Of course if one compactifies
the ambient transverse space, the 4d graviton reappears by the Kaluza-Klein
mechanism.

\section{Conformal gauge theories in higher dimensions}

Although our main goal is to find a string theoretic description of the
asymptotically free theories, conformal cases are not without interest. They
are easier and can be used as a testing ground for the new methods. In this
section we briefly discuss conformal bosonic gauge theories in various
dimensions [2 ]. The background in these cases is just the AdS space. We
have to perform the non-chiral GSO projection in order to eliminate the
boundary tachyon (it would add an instability to the field theory under
consideration; we do not consider here an interesting possibility that this
instability resolves in some new phase).

The GSO projection in the non-critical string is slightly unusual. Let us
consider first d=5 (corresponding to the d=4 gauge theory). In this case we
have 4 standard NSR fermions $\psi _{\mu }$ on the world sheet and also a
partner of the Liouville field $\psi _{5}$. For the former we can use the
standard spin fields defined by the OPE 
\begin{equation}
\psi _{\mu }\times \Sigma _{A}\sim \left( \gamma _{\mu }\right) _{AB}\Sigma
_{B}
\end{equation}

where we use the usual 4$\times 4$ Dirac matrices. The spinor $\Sigma _{A}$
can be split into spinors $\Sigma _{A}^{\overset{+}{-}}$ with positive and
negative chiralities. It is easy to check that the OPE for them have the
structure 
\begin{eqnarray}
\Sigma ^{\overset{+}{-}}\times \Sigma ^{\overset{+}{-}} &\sim &(\psi
)^{[even]} \\
\Sigma ^{\overset{+}{-}}\times \Sigma ^{\overset{-}{+}} &\sim &(\psi
)^{[odd]}
\end{eqnarray}

The symbols on the RHS mean the products of even/odd number of the NSR
fermions. Notice that this structure is the opposite to the one in 10
dimension. In order to obtain the spin operators in 5d we have to introduce
the Ising order and disorder operators , $\sigma $ and $\mu $ , related to $%
\psi _{5}$. These operators are non-holomorphic and correspond to RR-states.
Their OPE have the structure 
\begin{eqnarray}
\sigma \times \sigma &\sim &(\psi _{5})^{[even]} \\
\sigma \times \mu &\sim &(\psi _{5})^{[odd]}
\end{eqnarray}

Using these relations we obtain the following GSO projected RR spin operator 
\begin{equation}
\Sigma =( 
\begin{array}{cc}
\sigma \Sigma ^{+}\overline{\Sigma ^{+}} & \mu \Sigma ^{+}\overline{\Sigma
^{-}} \\ 
\mu \Sigma ^{-}\overline{\Sigma ^{+}} & \sigma \Sigma ^{-}\overline{\Sigma
^{-}}
\end{array}
)
\end{equation}

It has the property 
\begin{equation}
\Sigma \times \Sigma \sim \left( \psi \right) ^{[even]}
\end{equation}

needed for the non-chiral GSO projection, which consists of dropping all
operators with the odd number of fermions. Notice also that the non-chiral
picture changing operator has even number of fermions. The RR matrix $\Sigma 
$ has 16 elements. We can now write down the full string action in the AdS$%
_{5}$ space. It has the form 
\begin{equation}
S=S_{B}+S_{F}+S_{RR}+S_{ghost}
\end{equation}

where $S_{B}$ is given by (16), $S_{F}$ by 
\begin{equation}
S_{F}=\int d^{2}\xi \lbrack \overline{\psi }_{M}\nabla \psi _{M}+\frac{1}{%
\sqrt{\lambda }}(\overline{\psi _{M}}\gamma _{\mu }\psi _{N})^{2}]
\end{equation}

where $M=1,...5,$and one have to use the standard spin-connection projected
from AdS on the world sheet in the Dirac operator. \ So far we are
describing the usual action of the sigma model with N=1 supersymmetry on the
world sheet. The unusual part is the RR term given by 
\begin{equation}
S_{RR}=f\int d^{2}\xi Tr(\gamma _{5}\Sigma )e^{-\frac{\phi }{2}}
\end{equation}

Here $e^{-\frac{\phi }{2}}$ is a spin operator for the bosonic ghost [12 ]
and $f$ is a coupling constant (which \ in one loop approximation is equal
to 1).

I believe that this model can be exactly solved, although it has not been
done yet. A promising approach to this solution may be based on the
non-abelian bosonization [13] in which the fermions $\psi _{M}$ are replaced
by the orthogonal matrix $\Omega _{MN}$ with the WZNW Lagrangian. In this
case the RR term is simply a trace of this matrix in the spinor
representation. This formalism lies in the middle between the NSR and the
Green-Schwartz approaches and hopefully will be useful. Meanwhile we will
have to be content with the one loop estimates which are justified in some
special cases listed below and help to get a qualitative picture in general.
In this approach one begins with the effective action 
\begin{equation}
S=\int d^{d}x\sqrt{G}e^{\Phi }\left[ \frac{d-10}{2}-R-\left( \nabla \Phi
\right) ^{2}\right] +\int d^{d}xF_{d}^{2}\sqrt{G}
\end{equation}

Here $F_{d}$ is the RR d-form and $F_{d}^{2}=G^{A_{1}A_{1}^{^{\prime
}}}...G^{A_{d}A_{d}^{^{\prime }}}F_{A_{1}...A_{d}}F_{A_{1}^{^{\prime
}}...A_{d}^{^{\prime }}}$ ; the form $F_{d}$ will be assumed to be
proportional to the volume form . The dilaton field $\Phi $ is normalized so
that $e^{\Phi }=g_{s}^{-2}$, where $g_{s}$ is the string coupling constant.
Conformal cases involve either constant curvature solutions of the equations
of motion or the products of the manifolds with constant curvatures. The
dilaton in these cases is also a constant. Such an ansatz is very easy to
analyze. Let us begin with the single AdS$_{d}$ space. Consider the
variation of the metric which preserves the constancy of the curvature 
\begin{eqnarray}
\delta G_{AB} &=&\varepsilon G_{AB} \\
\delta R &=&\delta (G^{AB}R_{AB})=-\varepsilon R \\
\delta \Phi  &=&const
\end{eqnarray}

That immediately gives the relations 
\begin{eqnarray}
\frac{d-10}{2}-R &=&0 \\
e^{\Phi }(1-\frac{d}{2})R-\frac{d}{2}F_{d}^{2}=0 &&
\end{eqnarray}

If the assume that the flux of the RR field is equal to $N$, we get $%
F_{d}^{2}=N^{2.}$. If we introduce the coupling constant $\lambda
=g_{s}N=g_{YM}^{2}N,$ we obtain the background AdS solution \ [2 ] with 
\begin{equation}
\mid R\mid \sim \lambda ^{2}\sim 10-d
\end{equation}

We can trust this solution if $d=10-\epsilon $, in which case the curvatures
and the RR fields are small and the above one loop approximation is
justified. According to the discussion in the preceding sections, this
solution must describe the Yang-Mills theory, perhaps with one adjoint
scalar, in the space with dimension $9-\epsilon .$ We conclude that this
bosonic higher dimensional gauge theory has a conformally invariant fixed
point! It may be worth mentioning that this is not atypical for
non-renormalizable theories to have such fixed points. For example a
nonlinear sigma model in dimension higher than two, where it is
non-renormalizable, does have a conformal critical point at which the phase
transition to ferromagnetic phase takes place. However it is hard to say up
to what values of $\epsilon $ we can extrapolate this result.

These considerations allow for several generalizations. First of all we can
consider products of spaces with constant curvatures by the same method.
Take for example the space AdS$_{p}\times S_{q}$ with curvatures $R_{1}$and $%
R_{2}$ and $d=p+q$. To get the equations of motion in this case it is
sufficient to consider the variations 
\begin{eqnarray}
\delta G_{ab} &=&\varepsilon _{1}G_{ab} \\
\delta G_{ij} &=&\varepsilon _{2}G_{ij}
\end{eqnarray}

where the first part refers to AdS and the second to the sphere. A simple
calculation gives the equations 
\begin{eqnarray}
\frac{d-10}{2}-R_{1}-R_{2} &=&0 \\
&&(1-\frac{2}{p})R_{1}+\frac{10-d}{2}=-e^{-\Phi }F_{d}^{2} \\
(1-\frac{2}{q})R_{2}+\frac{10-d}{2} &=&e^{-\Phi }F_{d}^{2} \\
F_{d}^{2} &=&\lambda ^{2}R_{2}^{q}
\end{eqnarray}

Here we assume that the RR flux is permeating the AdS component of space
only, being given by the volume form. The last equation follows from the
normalization condition of this flux and the extra factor proportional to
the volume of $S_{q}$ in the action. Solving this equations we get 
\begin{eqnarray}
R_{1} &=&-(\frac{10-p-q}{2})\frac{p(q+2)}{p-q} \\
R_{2} &=&(\frac{10-p-q}{2})\frac{q(p+2)}{p-q}
\end{eqnarray}

This solution describes bosonic gauge theories with $q+1$ adjoint bosons;
again it can be trusted if the curvatures are small. Another generalization
is related to the fact that strictly speaking we must include the closed
string tachyon in our considerations. It was shown in [14 ] that there
exists an interesting mechanism for the tachyon condensation, following from
its couplings to the RR fields. It is easy to include the constant tachyon
field in our action and to show that it doesn't change our results in the
small curvature limit. According to [14], in the critical case $d=10$ the
tachyon leads to the running coupling constants. In the non-critical case
there is also a conformal option, described above, in which the tachyon
condenses to a constant value.

Finally let us describe the reasons to believe that the conformal solutions
can be extrapolated to non-small curvatures and thus the sigma model (42 )
has a conformal fixed point. The first two terms in (\ 45) are the expansion
of the sigma model central charge. When the couplings are not small, we have
to replace 
\begin{equation}
\frac{d-10}{2}-R\Rightarrow \frac{c(R)-10}{2}
\end{equation}

where $c(R)$ is the the central charge of the 2d sigma model with the target
space having \ a curvature $R$. It decreases , according to Zamolodchikov,
along the renormalization group trajectory. We can call this the second law
of the renormalization group. Let us conjecture that there is also a \emph{%
third law} 
\begin{eqnarray}
c\left( R\right) &\rightarrow &0\ \ (R\rightarrow +\infty ) \\
c\left( R\right) &\rightarrow &\infty \ \ (R\rightarrow -\infty )
\end{eqnarray}

The first property follows from the fact that usually the sigma models with
positive curvature develop a mass gap and thus there are no degrees of
freedom contributing to the central charge. The second equation is harder to
justify; we know only that $c(R)$ is increasing in the direction of negative
curvature.

With this properties and with some general form of the RR terms it is
possible to see that the conformal solution to the equations of motion,
obtained by the variations ( 46\ ), continue to exist when $\epsilon $ is
not small. Of course this is not a good way to explore these solutions.
Instead one must construct the conformal algebra for the sigma model
action(42 ). This has not been done yet.

\section{Infrared screening of the cosmological constant and other
speculations}

In this section we will discuss some speculative approaches to the problems
of vacuum energy and space-time singularities. I shall try to revive some
old ideas [15 ], adding some additional thoughts. The motivation to do that
comes from the remarkable recent observational findings indicating that the
cosmological constant is non- zero, and its scale is defined by the size of
the universe (meaning the Hubble constant). These results seem very natural
from the point of view advocated in [ 15], according to which there is an
almost complete screening of the cosmological constant due to the infrared
fluctuations of the gravitational field. This phenomenon is analogous to the
complete screening of electric charge in quantum electrodynamics found by
Landau, Abrikosov and Khalatnikov and nicknamed ''Moscow zero''. Here we
will try to argue in favor for another ''zero'' of this kind -that of the
cosmological constant.

Let us consider at first the Einstein action 
\begin{equation}
S=-\int \left( R-2\Lambda _{0}\right) \sqrt{g}d^{4}x
\end{equation}

Here $\Lambda _{0}$ is a bare cosmological constant which is assumed to be
defined by the Planck scale. It is clear from the form of the action that if
we consider the infrared fluctuations of the metric (with the wave length
much larger then the Planck scale) their interaction will be dominated by
the second term in this formula since it doesn't contain derivatives. To get
\ some qualitative understanding of the phenomenon, let us consider
conformally flat fluctuations of the metric 
\begin{equation}
g_{\mu \nu }=\varphi ^{2}\delta _{\mu \nu }
\end{equation}

The action takes the form 
\begin{equation}
S=-\int [\frac{1}{2}\left( \partial \varphi \right) ^{2}-\Lambda _{0}\varphi
^{4}]d^{4}x
\end{equation}

It has the well known feature of non-positivity. The way to treat it was
suggested in [16] and we will accept it, although it doesn't have good
physical justification. To use $S$ in the functional integral we will simply
analytically continue $\varphi \Rightarrow i\varphi ;$ after that the action
takes the form 
\begin{equation}
S=\int [\frac{1}{2}\left( \partial \varphi \right) ^{2}+\Lambda _{0}\varphi
^{4}]d^{4}x
\end{equation}

The infrared fluctuations of $\varphi $ are relevant and lead to the
screening of $\Lambda _{0}$%
\begin{equation}
\Lambda \sim \frac{1}{\log (M_{pl}L)}
\end{equation}

where $L$ is an infrared cut-off. More generally we could represent the
metric in the form 
\begin{equation}
g_{\mu \nu }=\varphi ^{2}h_{\mu \nu };\ \ \det (h_{\mu \nu })=1
\end{equation}

It is not known how to treat the unimodular part of the metric. We can only
hope that it will not undo the infrared screening although can change it.
Also the screening (67 ) with $L\sim H^{-1}$ (where $H$ is the Hubble
constant ) is not strong enough to explain the fact that $\Lambda \sim H^{2}.
$ \ It is not impossible that the two problems cure each other. When we have
several relevant degrees of freedom, the renormalization group equations
governing the $L$ dependence of various coupling constants including $%
\Lambda $ may have a powerlike asymptotic (in contrast with (67 )). Such
examples exist, starting from the cases with two independent coupling
constants. Thus the infrared limit \ of the Einstein action (perhaps with
the dilaton added ) may be described by a conformal field theory, giving the
cosmological constant\ defined by the Hubble scale. The renormalization
group should take us from Planck to Hubble.

Even if this fantasy is realized, we have to resolve another puzzle. It is
certainly unacceptable to have a large cosmological constant in the early
Universe, since it will damage the theory of nucleosynthesis. At the first
glance it seems to create a serious problem for the screening theory,because
when the universe is relatively small, the screening is \ small also and the
cosmological constant is large. The way out of this problem is to conjecture
that in the radiation dominated universe the infrared cut-off is provided by
the curvature of space-time, while in the matter-dominated era it begins to
depend on other quantities characterizing the size of the universe. If this
is the case, in the early universe we get the screening law $\Lambda \sim R$
instead of $\Lambda \sim H^{2}$ . Substituting it in the Einstein action we
find that at this stage the infrared mechanism simply renormalizes the
Newton constant and thus is unobservable. In the matter dominated era the
effective $\Lambda $ begin to depend on other things (like the Friedman warp
factor $a$ ) and that can easily give the observed acceleration of the
universe. Thus the change in the infrared screening must be related to a
trace of the energy momentum tensor. We can say that in the correct theory
the cosmological constant vanishes without the trace. To be more precise, it
is disguised as a Newton constant, until the trace of the energy-momentum
tensor reveals its true identity. The above picture have some remote
resemblance to the scenario suggested recently in [17].

In spite of the obvious gaps in these arguments, they give a very natural
way of relating the cosmological constant to the size of the universe and
thus are worth developing. Perhaps the AdS/CFT correspondence will be of
some use for this purpose. The main technical problem in testing these ideas
is the unusual $\varphi -$ dependent kinetic energy of the $h$-field.

We can also notice that the above scenario can explain the dimensionality of
space-time. Indeed, if this dimensionality is larger then four, the infrared
effects are small, the cosmological constant- large, and we end up in the
universe of the Planck size, which is not much fun.

This work was supported in part by the NSF grant PHY9802484.

\newpage

REFERENCES

[1] O. Aharony, S.Gubser, J. Maldacena, H. Ooguri, Phys. Rep. 323
(2000) 183

[2]  A. Polyakov J. Mod. Phys. A14 (1999) 645, hep-th/9809057

[3]  C. Callan et al. Nucl. Phys. B308 (1988) 221

[4]  A. Polyakov, V. Rychkov hep-th/0002106

[5] A. Polyakov, V. Rychkov hep-th/0005173

[6] N. Drukker, D. Gross, H. Ooguri Phys. Rev.D60 (199) 125, hep-th/9904191

[7] A. Sen JHEP 9912027, hep-th/9911116

[8] J. Polchinski Phys. Rev. Lett. 75 (1995) 4724, hep-th/9510017

[9] G. Horowitz, A. Strominger Nucl. Phys. B360 (1991) 197

[10] I. Affleck, A. Ludwig Phys.Rev.Lett. 67 (1991) 161

[11] L. Randall, R. Sundrum Phys. Rev. Lett. 83 (1999) 4690

[12] D. Friedan, E. Martinec, S. Shenker Nucl. Phys.B271 (1986) 93

[13] E. Witten Comm. Math. Phys.92 (1984) 455

[14] Ig. Klebanov A. Tseytlin Nucl. Phys.B547 (1999) hep-th/9812089

[15] A. Polyakov Sov. Phys. Uspekhi 25 (1982) 187

[16] G. Gibbons, S. Hawking Phys.Rev.D15 (1977) 2725

[17] C. Armendariz-Picon, V. Mukhanov, P. Steinhardt, astro-ph/0004139

\end{document}